# Characterizing Urban Lifestyle Signatures Using Motif Properties In Network of Places

Junwei Ma[1*], Bo Li[2], and Ali Mostafavi[3]

[1] (Corresponding author) Ph.D. Student, Urban Resilience.AI Lab, Zachry Department of Civil and Environmental Engineering, Texas A&M University, College Station, United States; e-mail: jwma@tamu.edu

[2] Ph.D. Student, Urban Resilience.AI Lab, Zachry Department of Civil and Environmental Engineering, Texas A&M University, College Station, United States; e-mail: libo@tamu.edu

[3] Associate Professor, Urban Resilience.AI Lab Zachry Department of Civil and Environmental Engineering, Texas A&M University, College Station, United States; e-mail: amostafavi@civil.tamu.edu

**Abstract:**

The lifestyles of urban dwellers could reveal important insights regarding the dynamics and complexity of cities. The availability of human movement data, anonymized and aggregated data captured from cell phone movement, enables characterization of distinct and recurrent human daily visitation patterns. Despite a growing body of research on analysis and evaluation of lifestyle patterns in cities, little is known about the characteristics of people's lifestyles patterns at urban scale. This limitation is primarily due to challenges in characterizing lifestyle patterns when human movement data is aggregated to protect the privacy of users. To address this gap, in this study, we model cities based on aggregated human visitation data to construct a network of places. We then examine the subgraph signatures (i.e., motifs) in the networks of places to map and characterize lifestyle patterns at city scale. Location-based data from Harris County, Dallas County, New York County, and Broward County in the United States were examined to reveal lifestyle signatures in cities. For the motif analysis, two-node, three-node, and four-node motifs without location attributes were extracted from human visitation networks. Second, homogenized nodes in motifs were encoded with location categories from NAICS codes. Multiple statistical measures, including network metrics and motif properties, were quantified to characterize lifestyle signatures. The results show that: i) people's lifestyles in urban environments can be well depicted and quantified based on distribution and attributes of motifs in networks of places; ii) motifs in networks of places show stability in quantity and distance as well as periodicity on weekends and weekdays indicating the stability of lifestyle patterns in cities; iii) human visitation networks and lifestyle patterns show similarities across different metropolitan areas implying the universality of lifestyle signatures across cities; iv) lifestyles represented by attributed motifs are spatially heterogeneous suggesting variations of lifestyle patterns within different population groups based on where they live in a city. The findings provide deeper insights into urban lifestyles signatures in urban studies and provide important insights for data-informed urban planning and management.

# 1. Introduction

Characterizing lifestyle patterns of people in cities is essential for understanding dynamics and complexity of cities arising from interactions between people, places, and activities (Batty, 2013). Understanding lifestyle signatures in cities could reveal important insights regarding ways people interact with their surrounding environment and places to inform urban planning decisions such as facility distribution, equity, sustainability, and resilience (Maeda et al., 2019; Toole et al., 2015).

Lifestyle can be defined based on sequence of life activities that residents in a city implement regularly over the course of a day. These life activities, in most cases, involve visitation to places (also called points of interest (POI)). Hence, with the growing availability of urban sensing and location-based data, a number of recent studies have examined human visitation and activities to unveil important characteristics about urban lifestyles (Xu et al., 2020; Zhang et al., 2021; Zhao et al., 2021). However, the existing literature related to characterizing population lifestyles is limited in two important aspects. First, the existing studies have examined the combination of visitation to places at the individual user level to unveil lifestyle patterns. Accessing and analyzing user-level location data is restricted due to privacy protection. Thus, the existing approaches are rather limited in terms of characterizing population lifestyles based on aggregated location-based data. Second, the approaches examine lifestyles based on sequence and grouping of visitations of places but do not account for spatial network structures embedded in urban lifestyle signatures. This limitation inhibits characterizing lifestyle patterns at urban scale while accounting for spatial network structures to reveal similarities and differences among lifestyles of residents across different cities.

In this study, we analyze motifs in network of places (i.e., human visitation network) for mapping and characterizing urban lifestyle signatures. The network of places represents nodes as places (points of interest) and links as visitation between nodes during a specific time period (e.g., day). The motifs are defined as common and recurrent subgraphs in the network of places. The networks of places for multiple US metropolitan areas (i.e., Harris County, Dallas County, New York County, and Broward County) are constructed using location-based data and points-of-interest data obtained from Spectus and SafeGraph. Next, two-node, three-node, and four-node motifs without location attributes were examined in the network of places. Finally, homogenized nodes in motifs were encoded with location categories from NAICS code. Statistical measures, including network metrics and motif properties, were quantified to characterize lifestyle signatures. Accordingly, the properties of lifestyle motifs, such as distribution, stability, and attributes were examined. The results demonstrate a multi-faceted portrait of population lifestyles at urban scale, which provides deeper insights into urban complexity and important implications for data-informed urban planning.

The remainder of this paper is organized as follows. Section 2 reviews related work in the area of human lifestyle characterization. Section 3 introduces the datasets that were utilized and describes the proposed methodological approach. Section 4 presents the results. Section 5 discusses the findings and concludes with contributions, policy suggestions and future work.

## 2. Background

The growing availability of urban-scale big data and sensing technologies has been a major driver of recent studies focusing on aspects of urban dynamics and complexity. We have recently witnessed the following becoming significant data sources for researchers to characterize urban dynamics and complexity: mobile phone calls from almost any location (Zhao et al., 2021), purchases made with credit cards (Xu et al., 2020), blog entries through online social media (Hasnat & Hasan, 2018), and even movements captured by video cameras (Zhang et al., 2021).

To obtain comprehensive pictures of spatial structure of cities, one important aspect of urban dynamics and complexity analysis is the characterization of population lifestyle patterns (Lazer et al., 2009). Human lifestyle analysis focuses on exploring spatial-temporal properties as well as hidden patterns behind the intra-urban and inter-urban movements (González et al., 2008). Some recent studies show how these lifestyle patterns (i.e., visitation to places) shape the spatial structure of cities. For example, Louail et al. used mobile phone data recorded in 31 Spanish cities to classify cities according to their commuting structure. They found that these cities essentially differ by their proportion of two types of movement flows: integrated (between residential and employment hotspots) and random flows, whose importance increases with city size (Louail et al., 2015).

Over the past several years, human lifestyle patterns have been growingly investigated. For example, daily activity patterns of students, part-time workers, retirees, telecommuters, drivers, women, and younger adults during the course of California workdays are recognized by using sequences of visits to different places in human mobility networks among different regions (Su et al., 2020). Varying topologies of trip combinations, such as "Home→Work→Post-work activity (for dining or shopping)→Back home", are extracted for transit riders from different public transportation systems in Nanjing, China based on temporal motifs (Lei et al., 2020). Despite the growing recognition of the importance of characterizing urban lifestyle pattern, the existing literature has limitations in two important aspects. First, the existing studies have examined lifestyle patterns using movement trajectories at the individual user level. The movement data at user level is not only difficult to access but also presents potential privacy issues. To overcome this limitation, aggregated data that are more privacy-compliant could be adopted in lifestyle pattern characterization. Second, a granular picture of human lifestyles cannot be obtained based on current approaches because most existing studies only focus on places of home, work and non-work. Urban areas are composed of different points of interest, such as restaurants, hospitals, and grocery stores, which contribute to the functionalities of urban locations and shaping of urban lifestyle signatures. Studying human lifestyle patterns by considering various points of interests is essential to provide more detailed insights into urban lifestyle signatures.

One approach to analyzing topological signatures in spatial networks (such as human visitation networks) is through examining motifs. Network motif is defined as the interconnection mode that has recurrence frequencies in the real network much higher than those in a randomized network (Dey et al., 2019). Milo et al. (2002) found that motifs ubiquitously exist in universal classes of networks, such as biochemical, neurobiological, ecological, and engineering network (Milo et al., 2002). As basic building blocks in networks, motifs are crucial for understanding the basic structures that control and modulate many complex system behaviors (Benson et al., 2016), which has been widely applied to network function studies in different disciplines in recent years (Stone et al., 2019). Recently, the concept of network motif has also attracted the attention of researchers in urban studies, who construct motifs arising from human movement trajectories, such as dockless bike-sharing ride recodes (Y. Yang et al., 2019) and public transportation card swipe records (Lei et al., 2020), to explore various human movement characteristics. These studies demonstrate the advantages of motifs in tapping into the universal human visitation patterns.

However, existing studies using motifs to characterize human lifestyle patterns lack the incorporation of node attributes, such as features of points of interest (e.g., restaurants, hospitals, and grocery stores). Since human visitations are shaped by the spatial configurations of facilities, lifestyles then are encoded in motifs of human visitation networks. Also, human visitations demonstrate uneven frequency of different locations, repeatedly returning to certain locations while being less likely to visit new ones (Song, Koren, et al., 2010; Song, Qu, et al., 2010). To this end, this study analyzed human visitation and location information datasets at an aggregated level, developed a method to differentiate categories of points of interest and extracted attributed motifs from human visitation network to characterize urban human lifestyle signatures.

### 3. Methodology

The methodological framework for examining urban lifestyle signatures is shown in Fig 1. This framework consists of four steps. The first step detects visits from point of interest to point of interest. The second step generates POI-to-POI networks, which we call the network of places. In the network of places, nodes are individual POIs and links are visitations between POI pairs during the same period of time (one day in this study). The weights of links capture the number of visits. For example, if ten users visit POI A and POI B during the same day, there would be a link between the two POIs with a weight value of ten. The construction of networks of places is essential for delineating lifestyle signatures in this study. The third step constructs human visitation motifs. The final step characterizes those human visitation motifs. These steps are explained in detail in the following subsections.

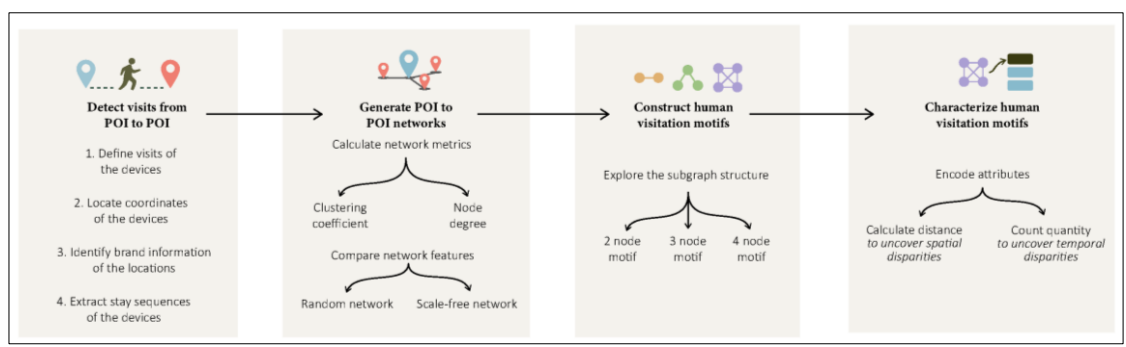

Fig 1 Workflow of the proposal analytical framework

## 3.1 Data sources

The data of this study came from two datasets. The first dataset is anonymized and privacy-enhanced mobile phone data provided by Spectus, an offline intelligence and measurement company using a location technology platform. Spectus provides high-quality location datasets by collecting first-party data from devices whose users opted in to the location data collection under a GDPR and CCPA compliant framework. The location data from Spectus is de-identified, and to further preserve privacy, the data provider removes sensitive points of interest from the dataset, and obfuscates home locations to the Census Block Group level. Researchers query Spectus data through an audited sandbox environment, and export aggregate datasets for research purposes. The dataset used in this study consists of anonymized location data in four metropolitan counties in the United States: Harris County and Dallas County in Texas, New York County in New York, and Broward County in Florida. The temporal range of the data is from February 1, 2020 through February 28, 2020. This period was selected since it was prior to the COVID-19 pandemic and its associated restrictions and thus could represent normal lifestyle activities. The data has a wide set of attributes, including anonymized device ID, POI ID, latitude, longitude, and dwell time of visitation. This dataset was used to detect POI visitation of devices during the studied period. The spatial coverage of POI data in this study can be seen in Fig 12 in Supplementary Information.

The location information of POIs was obtained from SafeGraph, Inc., a location intelligence data company that builds and maintains accurate POI data and store locations for the United States. The company maintains information of 6.5 million active POIs all over the United States. The dataset contains POI attributes, including POI ID, location name, address, category, and brand. All POIs are labelled with North American Industry Classification System (NAICS) category code, which is the standard used by Federal statistical agencies in classifying business establishments. The data structure in this study can be seen in Fig 13 in Supplementary Information.

## 3.2 Detecting visits from POI to POI

First, we used Spectus data to detect visits from starting POIs to destination POIs. Table "stop" from core data assets in Spectus was used to extract which POIs the devices have been to. Dwell time, which indicates stop duration of devices, was used as a criterion for defining a visit. If the duration an anonymized device stayed in one POI exceeded the thresholds, then a visit to that POI was recorded. Combing the time sequence of visits, the starting POIs and destination POIs were identified. In this way, a large number of visits from POI to POI were identified and aggregated for each week. Then POI ID was used to merge data between SafeGraph and Spectus datasets. In this way visits from POI to POI were affiliated with information of location categories. Finally, NAICS code was used to categorize the POIs. By doing so, the stay sequence of each individual device over a period of time can be extracted. Note that any device with only one stay in sequence was excluded because it indicated no movement through the whole day. These sequential stays will be employed to construct a graph in the next step. The outcome of this step includes aggregated data regarding the number of daily visits among POI pairs which is then used in constructing the network of places. Fig 2 illustrates the process of POI-to-POI visit detection.

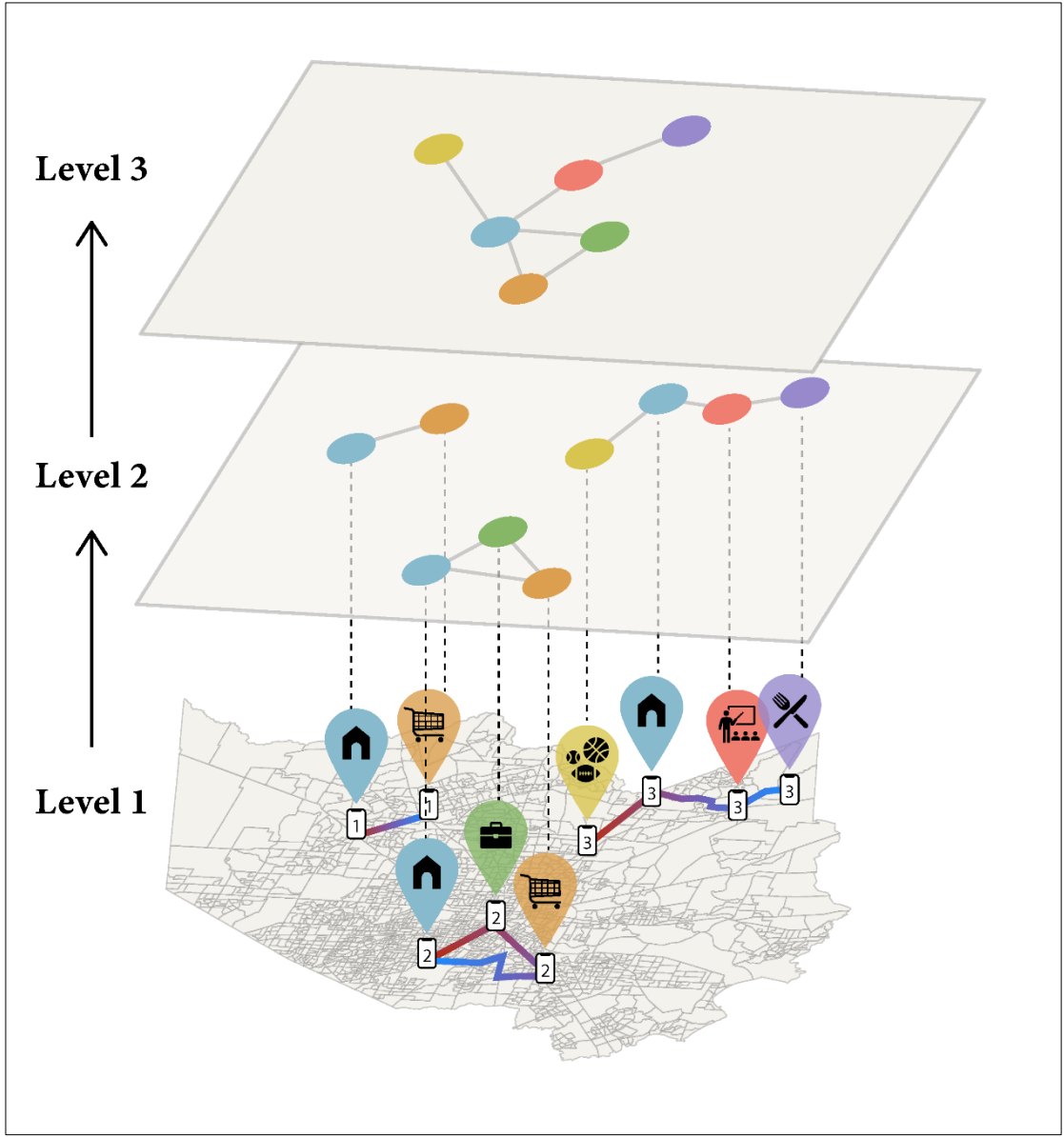

Fig 2 Illustration of POI-to-POI visits detection and network generation. Sequential stays represent a series of POIs where devices were engaged in activities during a day. Each stay is assigned a POI attribute from NAICS code (level 1). As a result, individual trajectories with different attributes are extracted (level 2). These sequential stays are used to construct a graph in level 3.

### 3.3 Generating POI-to-POI networks

To capture a global picture of urban human visitations, we aggregated all devices' trajectories to construct an undirected and weighted network that represented the sum of visit flows of all devices (level 3 in Fig 2). We named this network as the human visitation network (or network of places), which is defined as

$$G = (V, E, W) \qquad (1)$$

where, $V$ represents all of the distinct human visitation nodes (i.e., POIs) and $E$

represents all existing flows between pairs of nodes. The edge weights $W$ correspond to the counts of flows between two POI nodes. For each individual trip on edge, the weight is incremented by 1.

### 3.3.1 Network property metrics

In the first analysis step, we examined the general properties of the networks of places. Node degree and clustering coefficient are the most frequently used metrics to characterize properties of a network. Here, we employed these two metrics to characterize the properties in the networks of places.

**(1) Node degree $K_i$ and degree distribution $P(k)$**

Node degree $K_i$ and degree distribution $P(k)$ are important indicators that reveal the spatial heterogeneities of human movements (Saberi et al., 2018). In this study, nodes with larger degrees represent more highly connected places in the urban environment. The distribution of node degrees captures the number of nodes with a given degree $k$ in the global human visitation network. For a given network, node degree $K_i$ is defined as the number of nodes to which node is connected, as shown in Equation (2) (Scott & Carrington, 2011).

$$K_i = \sum_{i,j \in V} N(v_i, v_j) \tag{2}$$

where $(v_i, v_j)$ represents the edges in the network $G$.

**(2) Local and average clustering coefficient**

The local clustering coefficient of a node ($C_l$) is a measure of neighborhood density and captures the degree to which the neighbors of this node are linked with each other(Opsahl & Panzarasa, 2009). A high local clustering coefficient of a node indicates that devices who visit one POI will also frequently visit its neighbors in this study. For node $i$, its local clustering coefficient $C_l$ is the fraction of the links that are actually present among the total possible links between its neighbors. The equation for the local clustering coefficient of node $i$, as defined by (Barrat et al., 2004), which is

$$C_l(i) = \frac{1}{\sum_{i,j \in V} W(v_i, v_j) \cdot (k_i - 1)} \sum_{j,k} \frac{W(l_i, l_j) + W(l_j, l_k)}{2} a_{ij} a_{jk} a_{ki} \tag{3}$$

where $a_{ij}, a_{jk},$ and $a_{ki}$ are the elements of the adjacency matrix.

Then, the average clustering coefficient $C_a$ of all nodes in a network can be applied to quantify the density of the entire network.

$$C_a = \frac{\sum_{i \in V} C_l(i)}{N} \tag{4}$$

### 3.3.2 Reference networks

Reference networks are an important indicator measuring the possibility of the occurrence of certain network structures, given certain properties of empirical networks. In this study, two reference networks, which represent two extreme urban human visitation conditions that characterize urban spatial heterogeneity, were generated as follows.

**(1) Random human visitation network**

The random human visitation network represents entirely homogeneous neighborhoods in human visitation, which means that all devices' visitation flows are purely random, without any preferences. In this network, all resources and facilities in urban environment have a relatively uniform degree distribution and that individuals' visits are not restricted by the urban spatial structure. This network was simulated by means of random walks between any two nodes with the same number of nodes and probability as the global human visitation network in this study. The degree distribution of random human visitation network shows a Poisson distribution characteristic, which represents the property of homogeneity (Albert & Barabási, 2002). In addition, the average clustering coefficient of such network could be very small.

**(2) Scale-free human visitation network**

The scale-free human visitation network represents highly heterogeneous structures in the urban environment, indicating the spatial heterogeneity generated from the relative concentrations of POI facilities. Specific locations with more concentrated POIs attract a larger number of people, while other areas will experience minimal visits. The node distribution of this network follows a power law, which means most nodes have only a few edges, while a few nodes possess a large number of edges(Albert & Barabási, 2002). As will be shown in the results section, we examined the global properties in the network of places in comparison with random and scale-free reference networks.

**3.4 Constructing human visitation motifs**

In network theory, motifs are defined as common and recurrent subgraphs in networks (Dey et al., 2019). Accordingly, in this study, the human visitation motif denotes general urban lifestyle signatures. To identify patterns of population lifestyle signatures at urban scale, identifying and examining motifs in human visitation networks is a key step.

According to (Newman, 2018), motifs can be identified by exploring the network isomorphism. Let $G_1 = (V_1, E_1)$ and $G_2 = (V_2, E_2)$ be two graphs. If $V_2$ is a subset of $V_1$ ($V_2 \in V_1$) and $E_2$ is a subset of $E_1$ ($E_2 \in E_1$), then $G_2$ is a subgraph of $G_1$. There are two types of subgraphs, node-induced subgraphs and edge-induced subgraphs. A node-induced subgraph is a subset of the nodes of a graph G together with any edges whose endpoints are both in that subset, whereas an edge-induced subgraph is a subset of the edges of a graph G together with any nodes that are their endpoints. Since human visitation networks are node-based sequences, and node attributes play an essential role in differentiating lifestyles, we considered only node-induced subgraphs in this study.

Suppose there is a one-to-one mapping function $f: V(G_2) \rightarrow V(G_1)$, in which any two nodes $i$ and $j$ in $G_2$ are adjacent if and only if $f(i)$ and $f(j)$ are adjacent in $G_1$, then $G_1$ and $G_2$ are considered isomorphic ($G_2 \rightleftarrows G_1$). The mapping function $f$ is called an isomorphism between $G_1$ and $G_2$. When there is a subgraph $G_1'$ of $G_1$ ($G_1' \subset G_1$) and $G_1'$ is isomorphic to $G_2$, it means an appearance of $G_2$ in $G_1$. The total number of appearances is the frequency $F_G$ of $G_2$ in $G_1$. Once the frequency $F_G(G_2)$ exceeds a predefined cut-off value, $G_2$ is considered as a motif in $G_1$.

Motifs usually contains limited numbers of nodes, thus indicating the fundamental units to uncover the structural characteristics of a network (Milo et al., 2002). In this study, we scanned the human visitation network for all possible two-node, three-node, and four-node motifs. Finally, we identified one kind of two-node motif (M2-1), two kinds of three-node motifs (M3-1 and M3-2), and six kinds of four-node motifs (M4-1, M4-2, M4-3, M4-4, M4-5 and M4-6) in the global human visitation network. Table 1 summarizes basic structural characteristics of all motifs.

Table 1 Basic characteristics of motifs in global human visitation network

| Motif ID | Motif shape | Structural characteristic |
|---|---|---|
| M2-1 | 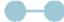 | Edge |
| M3-1 | 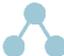 | Star, chain |
| M3-2 | 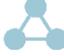 | Triangular, ring |
| M4-1 | 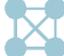 | Fully connected |
| M4-2 | 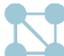 | 4-chordalcycle |
| M4-3 | 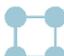 | Quadrangle, ring |
| M4-4 | 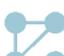 | 4-tailedtriangle |
| M4-5 | 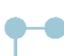 | Chain |
| M4-6 | 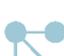 | Star |

### 3.5 Characterizing human visitation motifs

Characteristics of population lifestyles are not only reflected from the shape of motifs, but also the node attributes of motifs. For example, a two-node motif denoting visits from school to pharmacy is quite different from a motif denoting visits from grocery store to shopping mall. Therefore, considering motifs without node attributes could cause serious information loss for identifying unique urban lifestyle signatures. It is important to identify heterogeneous lifestyle signatures by affiliating motifs with detailed attributes (e.g., POI categories).

To classify location types of the nodes in network, this study adopted NAICS code to classify POI categories, which is a more accurate and direct way. The North American Industry Classification System code is developed for use by Federal Statistical Agencies for the collection, analysis and publication of statistical data related to the US economy. With the help of NAICS code, the primary business activities of a POI can be best depicted. The 2-digit NAICS codes and their business activities with visualized icons are shown in Table 2.

Table 2 Categories of business activities in 2-digit NAICS code

| NO. | Icon | NAICS code | Business activity |
|---|---|---|---|
| 1 | 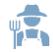 | 11 | Agriculture, Forestry, Fishing and Hunting |
| 2 | 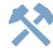 | 21 | Mining |
| 3 | 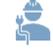 | 22 | Utilities |
| 4 | 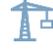 | 23 | Construction |
| 5 | 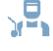 | 31, 32, 33 | Manufacturing |
| 6 | 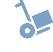 | 42 | Wholesale Trade |
| 7 | 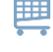 | 44, 45 | Retail Trade |
| 8 | 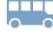 | 48, 49 | Transportation and Warehousing |
| 9 | 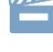 | 51 | Information |
| 10 | 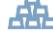 | 52 | Finance and Insurance |
| 11 | 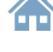 | 53 | Real Estate Rental and Leasing |
| 12 | 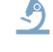 | 54 | Professional, Scientific, and Technical Services |
| 13 | 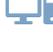 | 55 | Management of Companies and Enterprises |
| 14 | 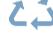 | 56 | Administrative and Support and Waste Management and Remediation Services |
| 15 | 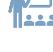 | 61 | Educational Services |
| 16 | 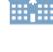 | 62 | Health Care and Social Assistance |
| 17 | 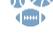 | 71 | Arts, Entertainment, and Recreation |
| 18 | 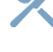 | 72 | Accommodation and Food Services |
| 19 | 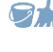 | 81 | Other Services (except Public Administration) |
| 20 | 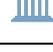 | 92 | Public Administration |

To enrich the human visitation motifs with NAICS attributes in a more specific manner, three scenarios were created (Fig 3). The left scenario is a two-node attributed visitation motif, the middle scenario is a three-node attributed visitation motif, and the right one is a four-node attributed visitation motif. The nodes in the motifs are physical locations (i.e., POIs), and the edges are the transitions representing people's movements among those locations.

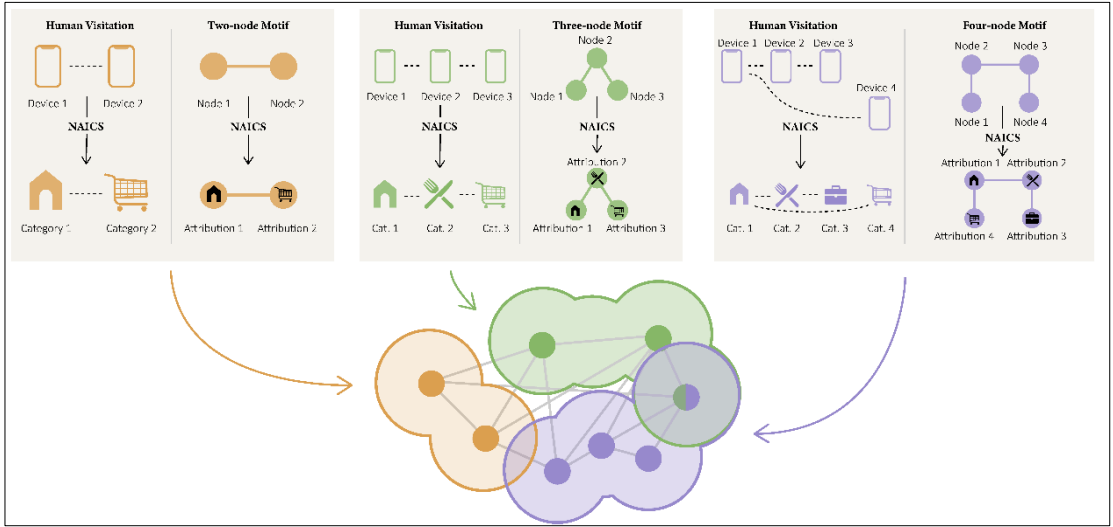

Fig 3 Human visitation motifs with attributes of two-digit NAICS code

Furthermore, lifestyles in urban environments can be measured from both spatial and temporal perspectives. Spatially, many scholars previously used spatial distance to measure a person's visitation pattern(Cao et al., 2021; Y. Yang et al., 2019; Yin & Chi, 2020). Accordingly, this study adopted average distance of motifs to measure the variability of lifestyles. Average distance of motifs is defined as the sum of the spatial length of each edge divided by the number of edges, where the spatial length of each edge (i.e. the linear distance between two POIs in reality) is calculated with reference to the Haversine Formula(Alam et al., 2016). The formula is as follows.

$$D = \frac{\sum_{i,j \in V} 2R \arcsin \sqrt{\sin^2\left(\frac{lat_i - lat_j}{2}\right) + \cos(lat_i)\cos(lat_j)\sin^2\left(\frac{lng_i - lng_j}{2}\right)}}{e} \quad (5)$$

Where, R is the radius of the Earth, $lat_i$, $lat_i$, $lat_i$, and $lat_i$ are the radian coordinates of node $i$ and node $j$, and $e$ is the number of edges of a motif.

Temporally, people's lifestyles vary over time. For example, many scholars have found that human movement behaviors during weekdays and weekends show large differences(McKenzie, 2020; Zhang et al., 2018). To ensure accuracy and avoid chance of error, this study uses the average number of attributed motifs on weekdays and weekends in a month to obtain more representative data to investigate lifestyle differences between weekdays and weekends.

## 4. Results
### 4.1 Properties of human visitation networks

We first investigated properties of the human visitation networks across the study areas. The networks were constructed from POI visit records of February 2020. Representations of the human visitation network for Harris County, Dallas County, New York County and Broward County are shown in Fig 4. For each network, properties we focused on were the number of nodes, edges, weight of edges, average degree and clustering coefficient (Table 3). Nodes represent POIs while the edges between the nodes represent the existing visits from the nodes they link. The weight of each edge is defined as the number of visits through the edge. The total weight of the network was calculated to show the number of the visits we adopted to construct the network. Harris County had the largest number of nodes and edges, while New York County has the fewest number of nodes, but the second largest number of edges, illustrating that human the visitation network in Harris County is relatively sparse, and that of New York County is denser. The total weights of the four counties are more than 1 million, indicating that the constructed networks are quite large scale.

Average degree is defined as the average number of links for each node(Scott & Carrington, 2011). New York County has a much higher average degree among the four counties, illustrating that POIs in New York County are interconnected more closely by human visits. Clustering coefficient measures the tendency to cluster within local nodes. All cluster coefficients in the four counties exceed 0.2, and the coefficient in New York County is 0.455, much higher than the average level. The relatively high clustering coefficients demonstrate that the human visitation networks are likely to shape local clusters, which supports the importance of motifs in characterizing lifestyle signatures at urban scale.

Table 3 Property metrics of global human visitation networks

| Study area (county) | Number of nodes | Number of edges | Weight of edges | Average degree | Clustering coefficient |
|---|---|---|---|---|---|
| Harris | 15931 | 136904 | 1735489 | 17.187 | 0.224 |
| Dallas | 9204 | 82090 | 1422144 | 17.838 | 0.248 |
| New York | 5312 | 104964 | 1232205 | 39.187 | 0.455 |
| Broward | 5713 | 72773 | 1365422 | 21.681 | 0.351 |

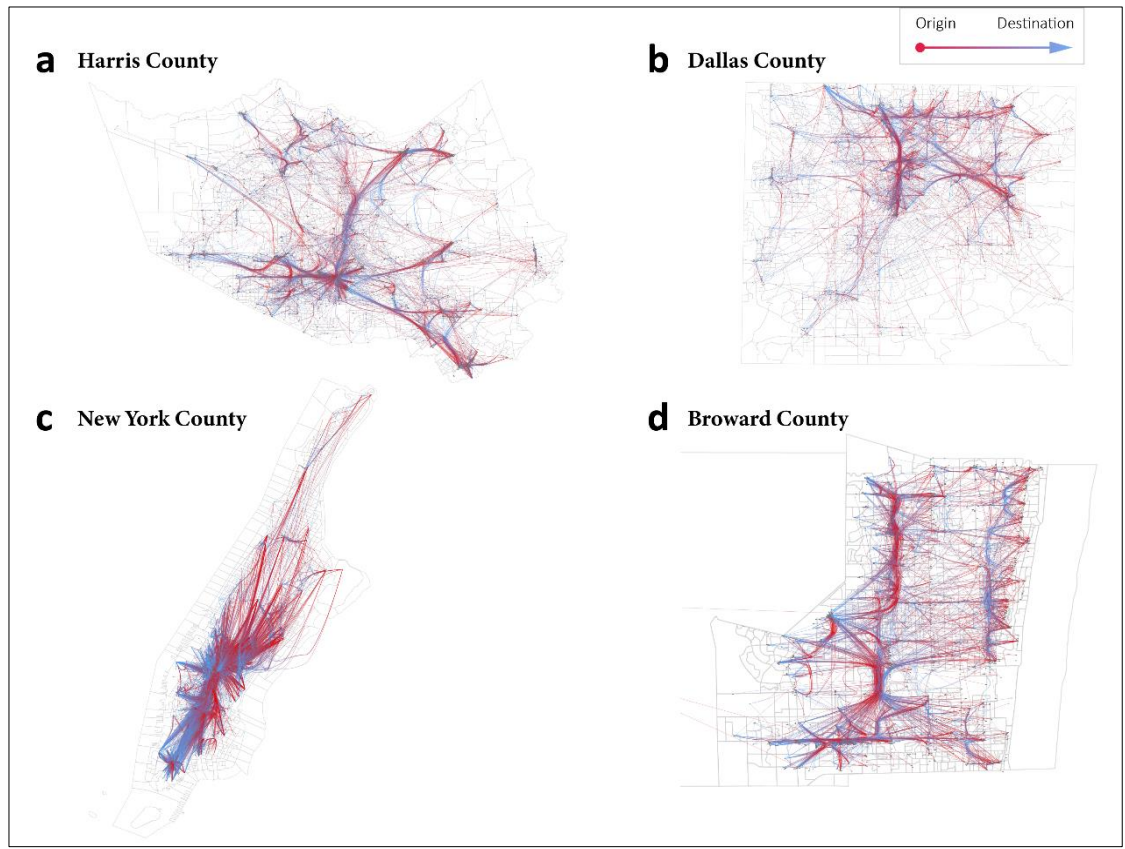

Fig 4 Geographic mapping of the human visitation networks

Next, we explored the visit frequency distribution of POIs in the global human visitation networks. Probability density function (PDF) (blue line) and its complementary cumulative density function (CCDF) (orange line) of the node degree were calculated as two metrics. PDF and CCDF of the four counties were plotted in Fig 5. Despite the different geographic and demographic characteristics across the four counties, the patterns of degree distribution are quite similar. There are steep decreases of possibility between the degree range from 1 to 10. The plots show that the number of nodes which have high degrees (greater than 100), is rather limited, while most nodes in the network have degrees less than 10. The phenomenon indicates the existence of node degree heterogeneity within the network of places. That is, large number of visits were heading to the very few POIs, while most POIs in the global human visitation network were visited just several times.

To further investigate the heterogeneity of the global human visitation networks, we calculated the Poisson distribution (random network) and power-law distribution (scale-free network), both of which were predicted with the same average degree as the original global human visitation networks. Fig 5 shows Poisson distribution (red line), and power-law distribution (green line) on a log-linear plot. PDF is observed to have a similar trend to power-law distribution, which could be fitted with an exponential distribution (the dashed line). The fat-tailed effect can be observed from the plots of PDF. A large set of visits are unique to individuals and rarely occur. In contrast, a small proportion of POIs are kept being visited by all the population. Such distribution of human visitations has been observed in all four counties in this study. PDF plots show serious deviation from Poisson distribution and similarity to power-law distribution, suggesting that the human visitation networks can be characterized as power-law distributions. Overall, the properties of the human visitation networks across the four countries take on similar patterns but vary in some extent. Within the counties, a small proportion of POIs in the human visitation network attracted a large number of visits, make it possible for us to explore the shared network substructures (namely motifs) within network.

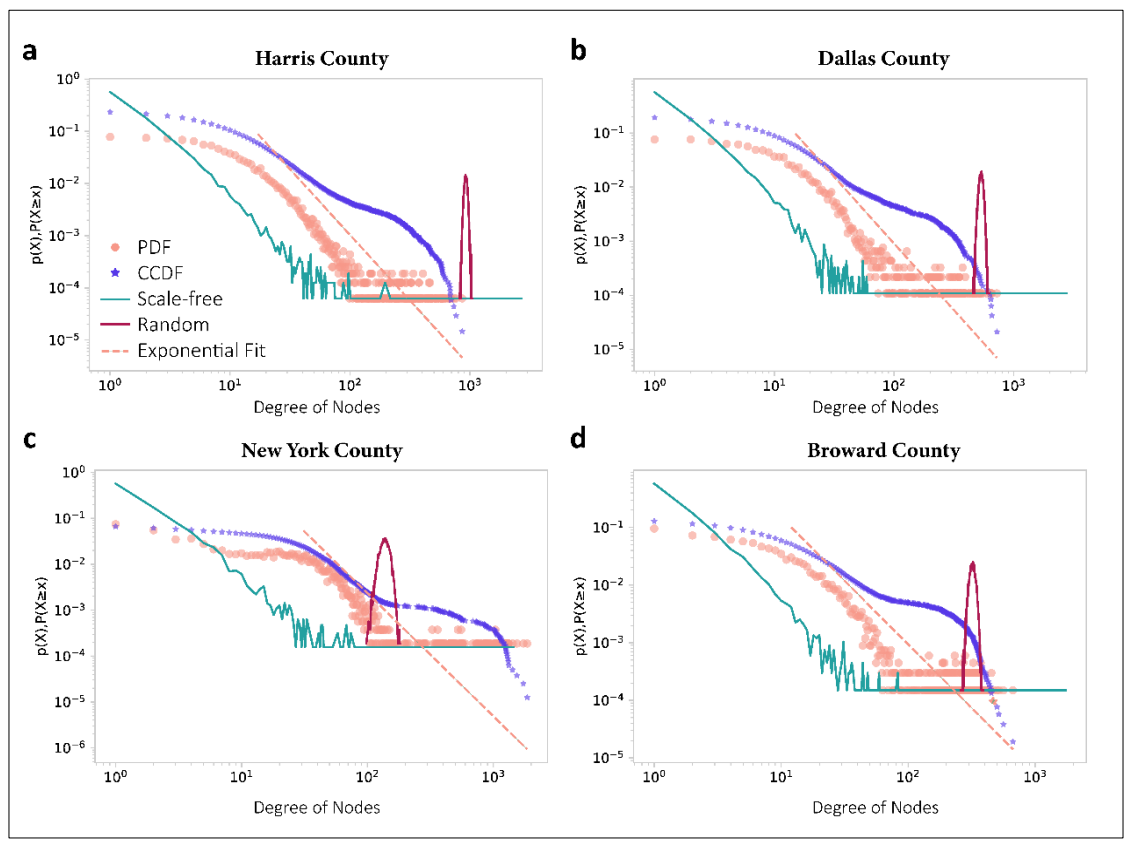

Fig 5 Distribution of node degree

**4.2 Properties of unattributed human visitation motifs**
Motifs are the basic substructures of a network, the examination of which could be examined to reveal the microscopic characteristics of the network. In this study, motifs in human visitation networks are a group of POIs which are linked by sequenced human visitation. Since motifs could capture the overlapped trajectory of human visits to places, we evaluate motifs to reveal urban lifestyle signatures shared by most population. In this study, we selected two-node, three-node, and four-node motifs, from human visitation networks of the four metropolitan countries. As shown in Table 4, nine types of motifs were extracted from the networks, representing different population lifestyles. Taking Harris County as an example, these nine motifs together accounted for 89.27% of the total number of motifs; the number of devices the motifs covered accounted for 76.66% of the total device count in the dataset; and the percentage of POI visits the motifs covered was 87.17%. The other three counties also have similarly high proportions.

Comparisons for proportions and average distances among the nine motifs were made to identify similarities and differences across the four counties as shown in Fig 6. The distributions of the two metrics are quite similar in the four counties. Four-node motifs have the highest proportion, indicating that people are more likely to go to four POIs. Among the categories of four-node POIs, it is worth noting that motifs M4-1 and M4-6 account only for very small percentages, probably because the visits represented by M4-1 are too complex, while the lifestyle represented by M4-6 is less efficient.

In terms of spatial distance, the average distance of M4-1 is the shortest among all motifs, and those of the rest of the motifs are roughly the same. Although M4-1 has the most edges, meaning that this lifestyle involves POI pairs that are visited altogether in a week; this type of lifestyle occurs primarily among POIs separated by a relatively small distance. In addition, although the distributions of motif distance across four counties are approximately the same, the average distance of every motif in New York County is significantly lower than that of the other three counties, as shown in spatial areas in Table 4. This result is probably because POIs are distributed more densely in New York County, so that people do not have to travel longer distances to reach each POI.

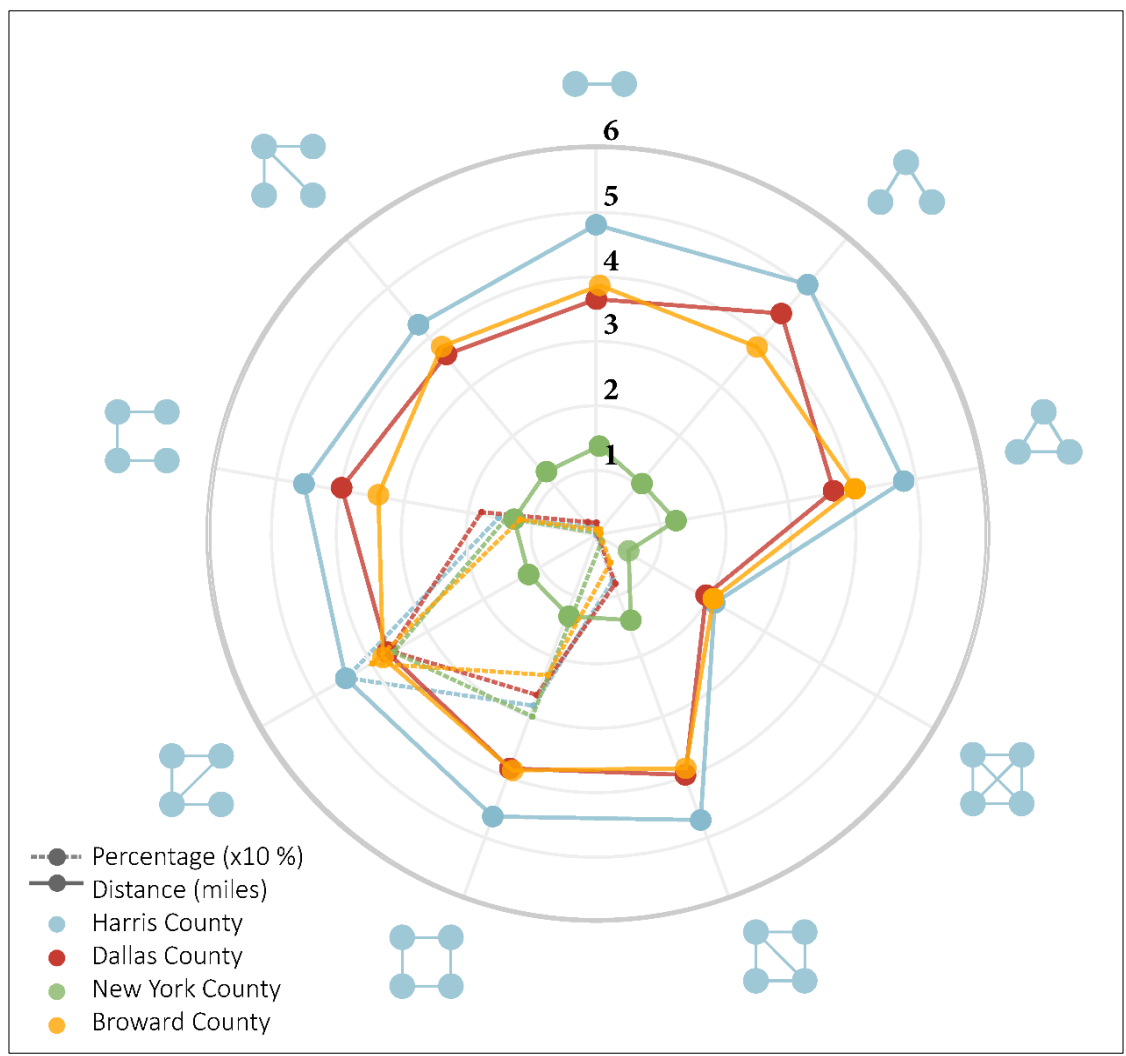

Fig 6 Distribution of number percentage and average distance in nine motifs

To uncover the temporal disparities in the nine categories of motifs, the distribution of daily percentage changes for nine motifs in one month are shown in Fig 7. Percentage changes were calculated based on a weekly pattern to differentiate weekdays and weekends. The orange line shows the change in the number of motifs per weekday over a 4-week period. The blue line reflects the change in the number of motifs per weekend over a four-week period. The purple line is the moving average curve plotted at 7-day intervals.

The moving average curves of the four counties show that the number of motifs is rather stable across all motif categories, despite minor fluctuations in some motifs (e.g., the number of M4-4 in Harris County fluctuates in the second and third week but stabilizes in the fourth week). This indicates that these motifs are stable enough over a considerable period of time to be adopted to depict urban population lifestyle signatures.

From a weekly perspective, both the number of motifs on weekdays and weekends follow a cyclical pattern across the four counties in general. For example, during weekdays, the number of M2-1 in Harris County almost always starts to drop on Monday, and drops to the lowest level on Tuesday, then rises quickly on Wednesday and Thursday, and finally drops slightly on Friday. On weekends, the number of M2-1 is always much lower on Sundays than on Saturdays. Except for M4-4, the rest of motifs basically have similar changes as M2-1, but the magnitude of the changes varies. In addition, the variation of percentage change of motif number on weekdays is significantly smaller than that on weekends. In Harris County, for example, most of the motifs vary within the range of 20% on weekdays, among which a significant proportion of the variations are even within the range of 10%. However, most of the motifs' change on weekends exceeds 20%. This indicates that lifestyles on weekdays are more stable, while more unstable on weekends.

To uncover the spatial disparities in these motifs, the distribution of the percentage changes in average distance for nine categories of motifs in one month are shown in Fig 8. Similarly, percentage change of average distance was also calculated in a weekly pattern. The orange and blue lines depict the change in distance for each weekday and weekend over a four-week period, respectively. The purple line is the moving average curve plotted at 7-day intervals. The moving average curves for the four counties are rather flat, showing that average distances of all categories of motif hardly change overtime. The percentage changes of all categories of motif are less than 5%, especially for M3-2, M4-1, M4-2, and M4-3, whose percentage changes in terms of average distance are close to zero. This result indicates that most people have fixed visitation patterns which produce stable lifestyle signatures at the urban scale. Obviously, the change in distance on weekends is also greater than that on weekdays. However, the tendency of the changes seems to be more chaotic, suggesting that people enjoy more flexible lifestyle patterns on weekends, such as choosing to go a farther place for shopping or dining. In general, the percentage change of motif number and average distance show that people's lifestyles can be well depicted and quantified by the nine categories of motif. Moreover, the two metrics are observed to be quite different within a week between weekday and weekends.

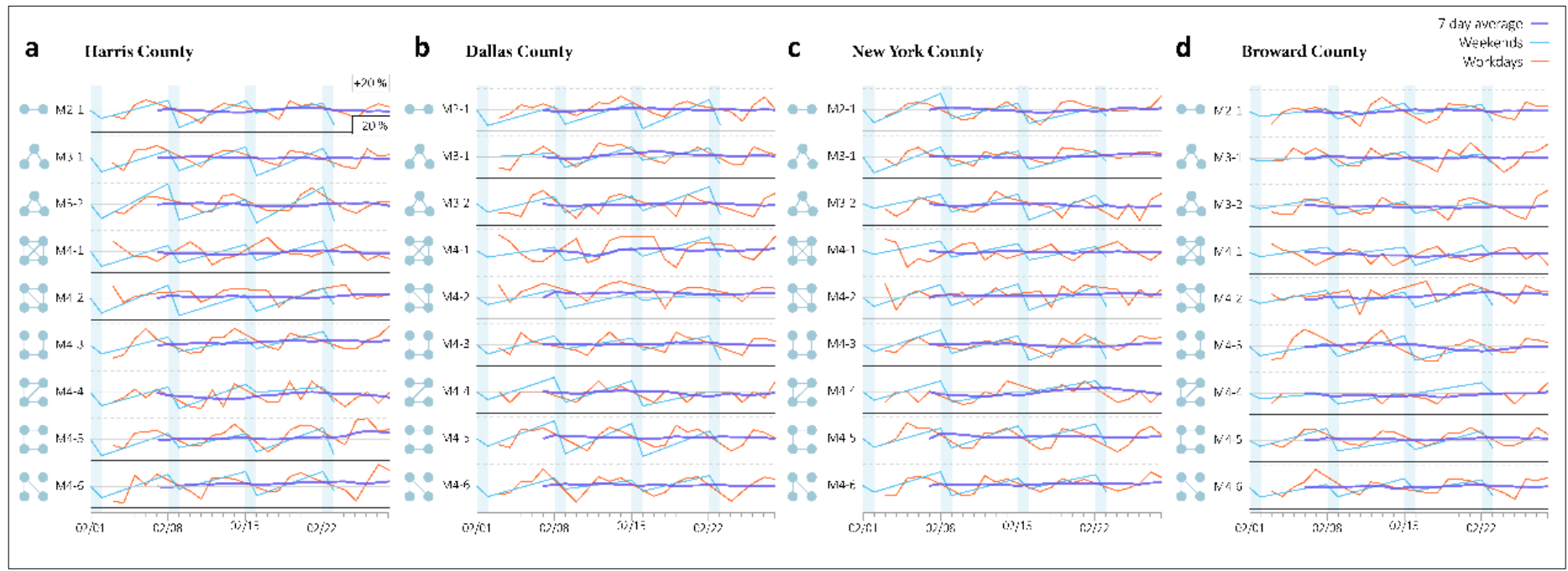

Fig 7 Daily distribution of percentage change of motif number

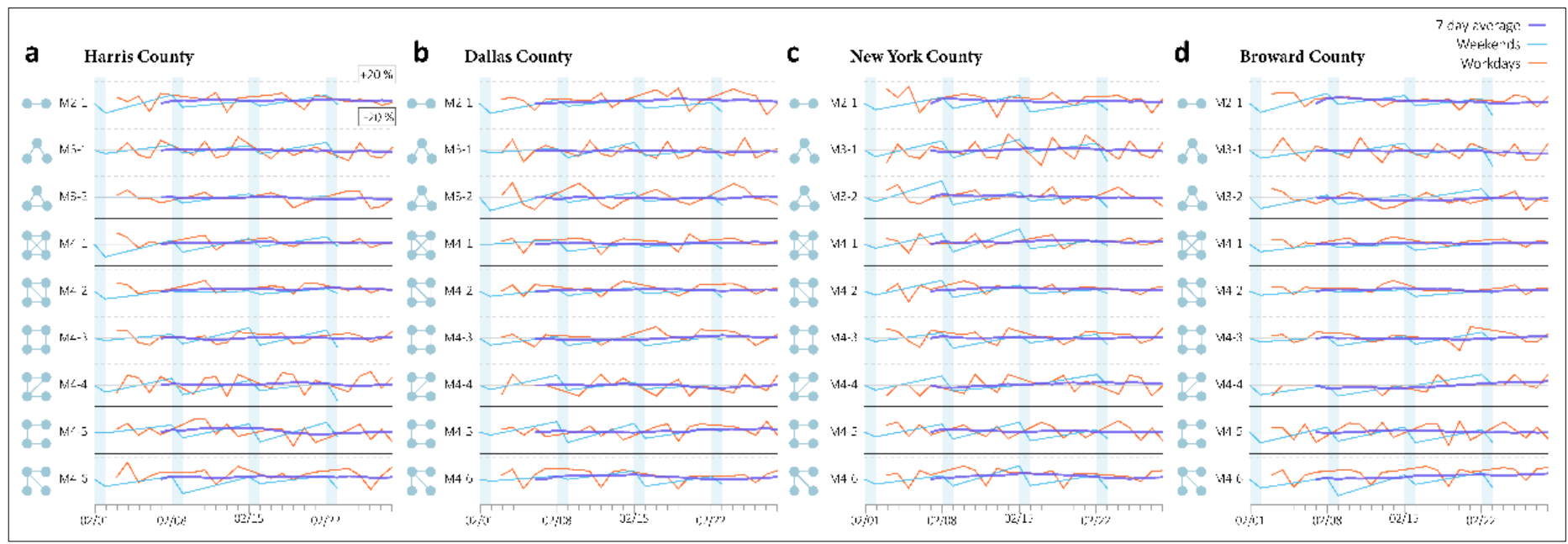

Fig 8 Daily distribution of percentage change of average distance

Table 4 Statistical properties of the motifs

| Study Area | Indicators | Global network | M2-1 | M3-1 | M3-2 | M4-1 | M4-2 | M4-3 | M4-4 | M4-5 | M4-6 |
|---|---|---|---|---|---|---|---|---|---|---|---|
| Harris County (4600 Km$^2$) | Number of motifs | 85237 | 1210 | 1441 | 13007 | 29304 | 24045 | 6418 | 34 | 256 | 375 |
| | Number of devices | 405562 | 5758 | 4532 | 40321 | 112209 | 114420 | 30532 | 152 | 1200 | 1776 |
| | Number of visit flows | 1735489 | 5758 | 9064 | 120963 | 673254 | 572100 | 122128 | 608 | 3600 | 5328 |
| | Percentage | / | 1.42 | 1.69 | 15.26 | 34.38 | 28.21 | 7.53 | 0.04 | 0.3 | 0.44 |
| | Distance | / | 4.811 | 4.256 | 4.569 | 4.452 | 4.651 | 4.709 | 2.109 | 4.812 | 5.068 |
| Dallas County (2350 Km$^2$) | Number of motifs | 66505 | 991 | 1377 | 11958 | 24427 | 17870 | 5640 | 20 | 200 | 279 |
| | Number of devices | 220703 | 3292 | 4563 | 39684 | 109752 | 40336 | 18708 | 76 | 640 | 928 |
| | Number of visit flows | 1422144 | 3292 | 9126 | 119052 | 658512 | 201680 | 74832 | 304 | 1920 | 2784 |
| | Percentage | / | 1.49 | 2.07 | 17.98 | 36.73 | 26.87 | 8.48 | 0.03 | 0.3 | 0.42 |
| | Distance | / | 3.615 | 3.597 | 3.985 | 3.721 | 3.901 | 4.007 | 1.95 | 3.709 | 4.429 |
| New York County (778.2 Km$^2$) | Number of motifs | 44968 | 85 | 81 | 6520 | 16454 | 13549 | 513 | 4 | 10 | 9 |
| | Number of devices | 329850 | 640 | 600 | 14829 | 120676 | 33408 | 3776 | 12 | 76 | 72 |
| | Number of visit flows | 1232205 | 640 | 1200 | 44487 | 724056 | 167040 | 15104 | 48 | 228 | 216 |
| | Percentage | / | 0.19 | 0.18 | 14.5 | 36.59 | 30.13 | 1.14 | 0.01 | 0.023 | 0.02 |
| | Distance | / | 1.365 | 1.265 | 1.333 | 1.256 | 1.358 | 1.423 | 0.522 | 1.201 | 1.023 |
| Broward County (3430 Km$^2$) | Number of motifs | 53222 | 335 | 436 | 6626 | 21480 | 12449 | 2571 | 5 | 53 | 80 |
| | Number of devices | 252414 | 1530 | 1986 | 30201 | 97872 | 56716 | 11704 | 28 | 228 | 364 |
| | Number of visit flows | 1365422 | 1530 | 3972 | 90603 | 587232 | 283580 | 46816 | 112 | 684 | 1092 |
| | Percentage | / | 0.63 | 0.82 | 12.45 | 40.36 | 23.39 | 4.83 | 0.01 | 0.1 | 0.15 |
| | Distance | / | 3.847 | 3.782 | 3.461 | 3.853 | 3.914 | 3.878 | 2.023 | 3.994 | 3.773 |

Note: Distances are measured in kilometers.

### 4.3 Properties of attributed human visitation motifs

Although motifs identified from the human visitation networks have revealed the temporal and spatial characteristics of urban lifestyle signatures from a topological perspective, we also explore the lifestyle heterogeneity by differentiating the motifs based on node attributes. As nodes represent POIs in human visitations, their attributes refer to the category information of POIs. In this study, we adopted NAICS code, which is one of the most commonly recognized standards to categorize these POIs. Motifs were differentiated according to node attributes so that they could reflect heterogeneous urban lifestyles.

The node attributes in motifs cover 20 categories of POIs by 2-digit NAICS code. Fig 9 (a) shows the frequency of visits for all categories, which show similar patterns across the four counties. POIs of the category of retail trade were the most frequently visited POIs, the percentage of which exceeds 22%, and even much higher in Harris County, Dallas County, and Broward County. Then, POIs in category of accommodation and food services ranked second, followed by other services.

To have a finer-grained understanding of urban lifestyle signatures, the top three ranked categories of POIs were selected and further specified into subcategories by 4-digit NAICS code. The ranked frequency of visits to POIs in the subcategories are shown on Fig 9 (b, c, d, e). Restaurants and other eating places take the lead with more than 20% of the visits across all counties. Personal care services and health and personal care stores are in the second or third place in Harris County, Dallas County, and Broward County. The commodities and services provided by POIs in these subcategories are closely related to people's basic needs of life, which explains the reason why these POIs are nearly the most frequently visited. Gasoline stations, grocery stores, and automotive repair and maintenance occupy the top five positions in these three counties, illustrating the importance of automobile usage and food consumption in people's daily life in those metro areas.

The distribution of visiting frequency to POIs in New York County is different from those in the other three counties. Clothing stores surprisingly ranked third and drinking places (alcoholic beverages) ranked fifth, while gasoline stations, grocery stores and automotive repair and maintenance, which ranked higher in other counties, ranked lower in New York County. As a dense metropolitan area, New York's bustling commercial facilities and convenient public transportation system have changed people's lifestyles to a certain extent. People (and visitors) in New York County have a much greater preference for fashion and entertainment compared to people from other counties. Likely owing to the highly developed public transportation network and denser distribution of POIs in New York County, residents are less reliant on automobiles.

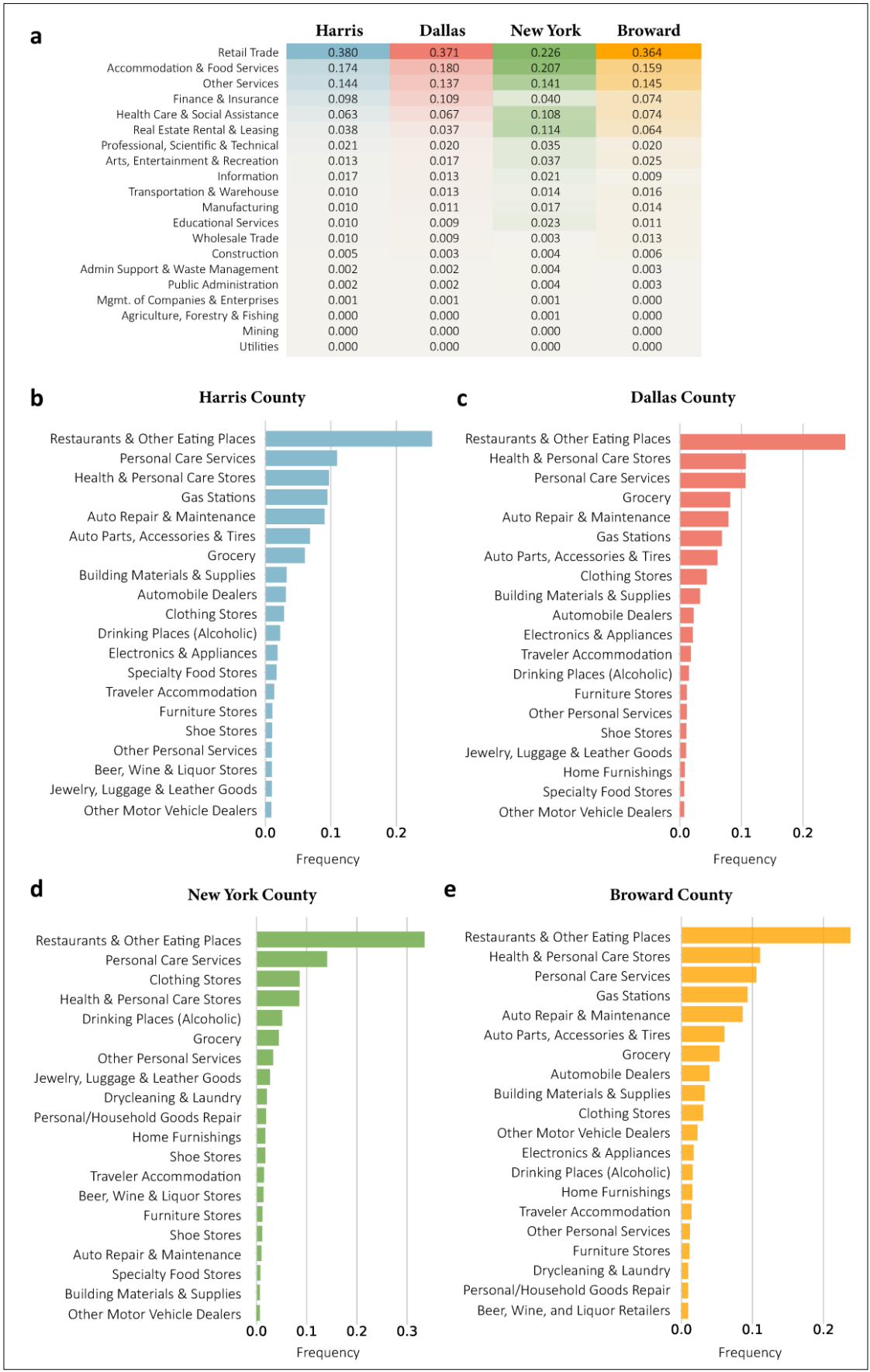

Fig 9 Distribution of visiting frequencies of POIs categories: (A) Frequency distribution of visits to POIs using 2-digit NAICS code; (b, c, d, e) Frequency distribution of visits to POIs using 4-digit NAICS code in Harris County, Dallas County, New York County, Broward County.

We have identified nine categories of motifs to capture typical urban lifestyle signatures. However, lifestyles represented by the same category of motifs may still be heterogeneous because the visited POIs may belong to various categories. Thus, we adopted a POI category represented by NAICS code as a node attribute to build an attributed motif. For each category of motifs, we selected the top 10 attributed motifs ranked by frequency in the four metropolitan counties, as shown in Fig 10.

For 2-node attributed motifs, the lifestyle services of retail trade—accommodation & food services and retail trade—other services are among the top four across the four counties, indicating that these services satisfy basic life needs. An interesting finding is that lifestyle retail trade—retail trade ranked top three in the other three counties but not for New York County. The phenomenon that retail trade—retail trade is less popular may provide insight into a new trend of lifestyle changes. Although purchasing commodities are essential for all the population, visiting stores may not be the only way. With more developed online shopping and logistic systems, population in New York County could order online and have commodities delivered to their houses. Lifestyle accommodation & food services— real estate rental & leasing ranked third in New York County, indicating the importance of residence, which may be a major concern in a highly urbanized area.

For 3-node attributed motifs, the most frequently present lifestyles are still those encompassing visits to retail trade, accommodation & food services, and other services in Harris County, Dallas County, and Broward County. The top three lifestyles in these counties encompass around 20% of the category of M3-1 and more than 10% in the category of M3-2, illustrating lifestyles in these areas have a consistent pattern. While in New York County, the top three lifestyles account only 6% and 8%, suggesting a more heterogenous distribution of lifestyles. This finding reveals that there is no dominant lifestyle in New York County, population, which adopts more heterogeneous lifestyles.

For 4-node attributed motifs, the top-ranked motifs show consistent patterns regardless of motif structures in all four counties. These motifs share nodes with the same attributes, suggesting that people may follow different trajectories but visit roughly the same categories of POIs. The identical categories of POIs and various co-visiting patterns show both the commonality and heterogeneity of people's lifestyles. Compared with 2-node and 3-node attributed motifs, the 4-node motifs have three distinctive features. First, POIs of retail trade have higher frequency to be involved in the top ranked motifs, illustrating that commodity purchase plays an even more important role as lifestyle complexity increases. Second, the percentage of various 4-node attributed motifs are more even, showing more diversity in urban lifestyles. Third, POIs with the categories of health care & social assistance, finance & insurance, educational services, and real estate rental & leasing appear more often in 4-node attributed motifs, which implies significant dimensions of urban life, such as residence, finance, health care and education for certain population subgroups.

Across all attributed motifs, there exist visits whose starting point and destination point are within the same category of POI, indicated as a circle in Fig 10. Nodes attributes of such motifs can be retail trade, accommodation & food services, and other services. Further examination of visitation data shows that although the visited POIs belong to the same category in 2-digit NAICS code, they may locate in different subcategories of 4-digit NAICS code. This finding indicates that there exists a lifestyle in which people tend to categorize and finish similar tasks, such as shopping, within one day. More analysis is needed to reveal this lifestyle characteristics by taking subcategories of POIs into account (which is beyond the scope of the current study).

Since POI data adopted in this study are affiliated with geo-information, we measured actual spatial distance of attributed motifs. Average distance during weekdays and weekends was calculated respectively to explore potential temporal patterns. Attributed motifs whose average distance ranked top 20 are shown in Fig 11. The disparity of urban lifestyles regarding average distance can be examined from two aspects. First, the most obvious observation is that average distance on weekends is longer than that on weekdays, which is consistent with the common sense that people tend to travel a longer distance during weekends because of fewer time constraints they may be bound by on weekdays. Second, attributed motifs ranked in the top three regarding average distance are quite different between weekdays and weekends. On weekends, the dominant categories of POIs are retail trade and accommodation & food services. On weekdays, in contrast, although these two categories also occupy a major proportion, visits to other categories of POIs, including health care & social assistance, finance & insurance, and other services are noted. The greater diversity of POI categories shown on weekdays may be explained by multiple reasons. For example, people visit certain POIs only during weekends. POIs in the aforementioned categories usually have longer business hours on weekends, which could also facilitate more visits.

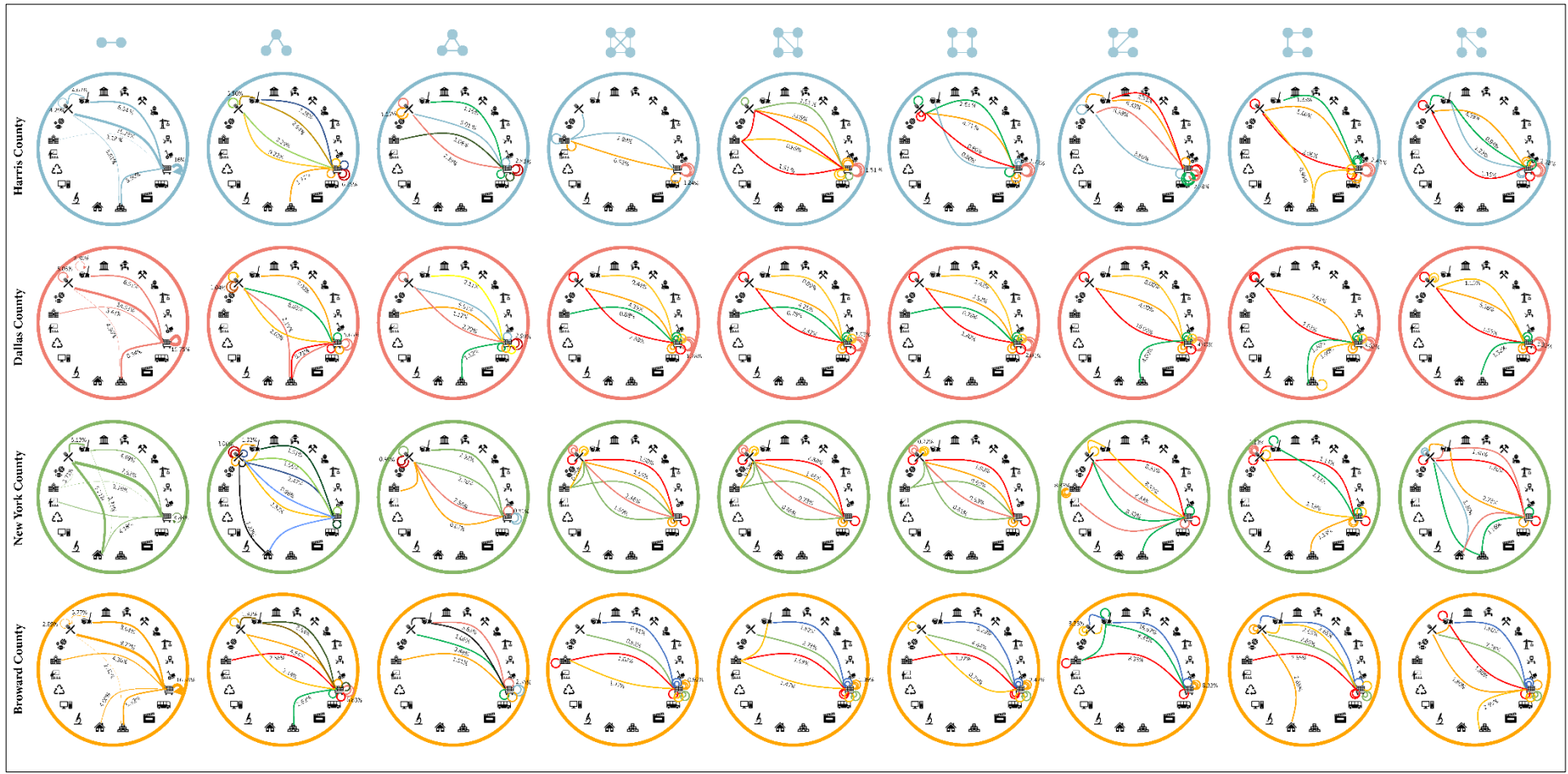

Fig 10 Frequency distribution of attributed motifs: Icons represent POI categories, among which links of identical color represent an attributed motif, namely a lifestyle. The frequency percentage of each lifestyle was calculated and marked next to the link.

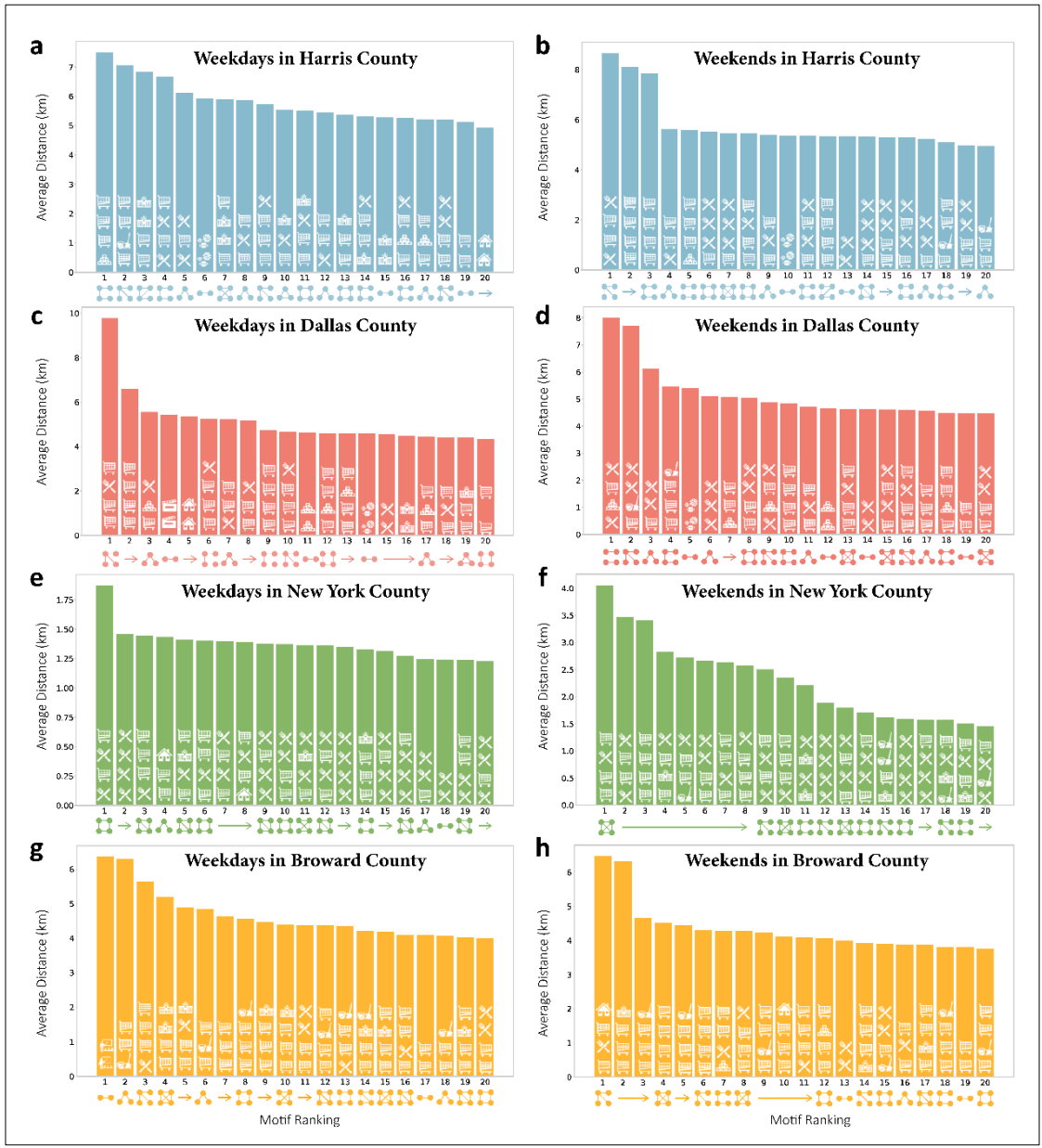

Fig 11 Distribution of average distance in attributed motifs Each bar is labeled with motif ID and icons representing categories of each node. The left part shows the ranking of average distance for the four counties on weekdays, and the right part shows the ranking on weekends.

## 5. Discussion and Conclusion

The overarching goal of this study is to gain insights into detailed human lifestyle signatures in an urban environment. Lifestyles are complex and vary among individuals. Existing studies using human mobility data to infer people's daily activities have suggested that lifestyles are repetitive, regular, and cyclical. To further characterize urban lifestyle signatures, this paper evaluated motifs in human visitation networks (network of places). In contrast to existing metrics that focus on the aggregation of human visitations, this paper constructs multiple human visitation subgraphs (motifs) extracted from human visitation networks to represent differentiated lifestyles in multiple metropolitan areas. Considering Harris County, Dallas County, New York County, and Broward County in the United States as cases, we have experimentally demonstrated a multi-facet portrait of urban lifestyles.

The statistical measures of the human visitation networks reveal macroscopic characteristics of human visitation to places. The results show that the disparity between numbers of points of interest and visits results in a high average degree and clustering coefficient. This tendency, to some extent, contributes to the formation of motifs that represent the prevalent visitation structure. Further, by examining the probability density function, cumulative density function, and complementary cumulative density function of the network, we find that the node degree approximates power-law distribution. This finding demonstrates that the greatest number of human visitations are concentrated in a small set of POIs, while some other POIs are rarely visited, which is consistent with previous studies on urban mobility (Di Clemente et al., 2018; Lenormand et al., 2015). Moreover, all four counties in this study exhibit similarity of their human visitation networks, demonstrating consistency in the general structure of human visitations, despite the significant differences in socio-demographic dimension and spatial structures. This study provides additional insight to this theory by drawing similarities between cross-sectional comparisons among different counties in United States, similarly to previous studies, such as those on Chicago (C. Schneider et al., 2013), Boston (Yin & Chi, 2020), San Francisco (Li et al., 2021), Singapore (Jiang et al., 2017), and Shenzhen (Cao et al., 2021).

By investigating the properties of unattributed human visitation motifs, the analysis results demonstrates that people's lifestyles show long-term stability in both motif quantity and distance and exhibit different patterns periodic recurrence on weekdays and weekends. The nine motifs proposed in this study can well depict and quantify the urban lifestyles of most populations. The results indicate the stability of lifestyle signatures during weekdays and less stability during weekend. This result is consistent with the insight that individuals show strong regularities of movements that tend to follow certain typical motifs, as reported in previous studies (Cao et al., 2019; Di Clemente et al., 2018; C. Schneider et al., 2013; C. M. Schneider et al., 2013). Moreover, people are more inclined to conduct lifestyles in a manner of 4-node structures. We conjecture that one major reason for this finding is that such lifestyle is more efficient and have easier access to resources. The result is in line with some empirical studies on the human activities in metropolitan cities, which find that residents in urban areas have simple but settled daily activity routines (X. Yang et al., 2019; Yin & Chi, 2020). This finding can be further explained by the possible determination of the abundance level of urban resources (i.e., bus stations, shopping malls, hospitals, etc.). Population with more abundant urban resources and higher-level socioeconomic level may have more efficient lifestyles.

Finally, the exploration of the properties of attributed motifs provides insights to the spatial and temporal heterogeneity among different lifestyles. From the spatial perspective, this study finds that people visit a variety of different POIs. Previous explorations of lifestyles have focused solely on a few types of visitation patterns, such as home, work, and non-work. (Huang & Wong, 2016; Jurdak et al., 2015; Xu et al., 2015), and there is no clear depiction of the complete pictures of lifestyle signatures embedded in different categories of locations. This study not only constructs a more complete mapping of diverse urban lifestyle patterns, but also reveals the frequently visited POIs and visited POIs represented by attributed motifs. From the temporal perspective, this study showed distinctions in different temporal lifestyles between weekdays and weekends using attributed motif quantity and average distance, which is a complement to the study of urban morphology and structure. In addition, it is worth noting that although, in general, the four counties in our study are similar in terms of lifestyle patterns, there are still some differences in the case of New York County. These differences, according to the analysis in this study, are not unrelated to the socio-demographic dimensions within this county. Previous studies have also suggested the need for more integration of infrastructure (e.g., public transportation) and geographic features of society and economy, such as income, race, wealth, or ethnicity, when conducting lifestyle interpretation in one region (Moro et al., 2021), which motivates us to provide more insights into diversified lifestyles in the future. The approach presented in this paper provides a new way to objectively distinguish lifestyle patterns across cities at a finer scale than what has been shown before.

The findings obtained in this study have multiple theoretical contributions. Understanding human lifestyle patterns has been a fundamental problem in urban science and city planning. While the universal laws and predictability of lifestyle patterns have been unveiled in past studies, the interaction among population and locations enabled by access to anonymized, aggregated databases, is still an area of active study and promises further insights into populations and their lifestyles. First, this study describes that people's lifestyles follow specific and regular patterns at the city level and exhibit temporal, spatial, and structural heterogeneity within counties, which expands our understanding of capturing human lifestyle patterns and exploring many other urban phenomena associated with human lifestyles. Second, this study exhibits the power-law nature of visitations to urban facilities. The insight provides a deeper understanding of urban structure, which thus can help policymakers evaluate their urban development strategies, especially urban resources allocation and city planning. Last, this study reveals human lifestyle signatures by considering characteristics of attributed motifs based on network of places using intelligent location-based data, which provides more data-driven and complexity-informed perspectives to understanding of the dynamics of populations and places in cities.

In addition, our work provides important implications for urban planning and development. On the one hand, this study advances the understanding of the stability and regularity of urban lifestyles that interact with different locations, which allows us to focus more on population dynamics in cities beyond the standard origin-destination studies of human mobility. Therefore, insights revealed by evaluation of this type of data would be important to inform urban planners when developing and proposing appropriate planning policies, including redistributing existing facilities and developing new facilities, on the premise of meeting the needs of urban residents to meet the lifestyle patterns of people. On the other hand, due to the power law effect in the network of places, locations with a large number of visitations will attract more visits, while those with fewer visitations will continue to remain in the same state. Therefore, the role of facility accessibility in these locations is central to improving urban and transportation planning. Also, the approach presented in this study provides a data-driven and quantitative way to compare different cities and evaluate the relationship between lifestyle signatures and city-level measures such as energy usage, equity, and access.

This work also has couple of limitations, which could be addressed in the future. First, lifestyle patterns may vary among people with different social-demographic attributes, such as age, gender, race and income. Quantifying the relationship between social-demographic characteristics and lifestyle patterns contributes to the understanding of social variations of urban structures. In our datasets, however, we cannot obtain social-demographic information about individuals due to privacy protection concerns. Future studies could find ways to integrate our anonymized location-based data with other datasets to examine the influence of social-demographic features on human lifestyles. Second, this study mainly looks into the similarities of lifestyle patterns across various counties. It should be noted that, although our case studies verify the similarity among counties from disparate regions of the United States, the sample size is still small and concentrates in metropolitan-type counties. One limitation in increasing the number of cities is the time and computational resources needed for data processing. Future studies should expand the sample counties based on factors such as population, urban typologies, and road networks to unravel the disparities of human lifestyles in different counties in association with other city characteristics to improve the understanding of how city characteristics shape lifestyle signatures.

**Acknowledgement**

This material is based in part upon work supported by the National Science Foundation under Grant CMMI-1846069 (CAREER), the Texas A&M University X-Grant 699, and the Microsoft Azure AI for Public Health Grant. The authors also would like to acknowledge the data support from Spectus, Inc. Any opinions, findings, conclusions, or recommendations expressed in this material are those of the authors and do not necessarily reflect the views of the National Science Foundation, Texas A&M University, Microsoft Azure or Spectus Inc.

**Data availability**

All data were collected through a CCPA- and GDPR-compliant framework and utilized for research purposes. The data that support the findings of this study are available from Spectus and SafeGraph, but restrictions apply to the availability of these data, which were used under license for the current study. The data can be accessed upon request submitted on cuebiq.com. Other data we used in this study are all publicly available.

**Code availability**

The code that supports the findings of this study is available from the corresponding author upon request.

**Supplementary Information**

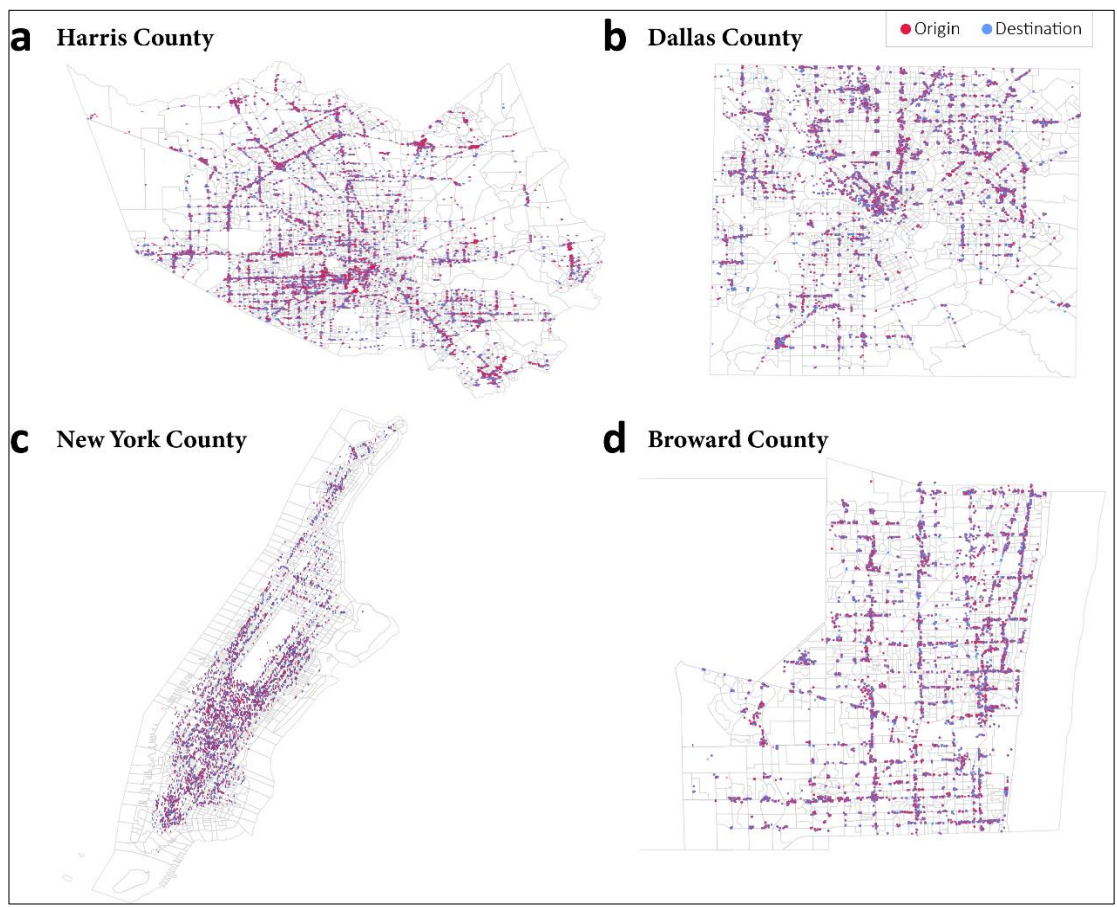

Fig12 Spatial coverage of POI data in this study. The red nodes in each graph represent the starting POIs of visitations, and the blue nodes represent the destination POIs of visitations

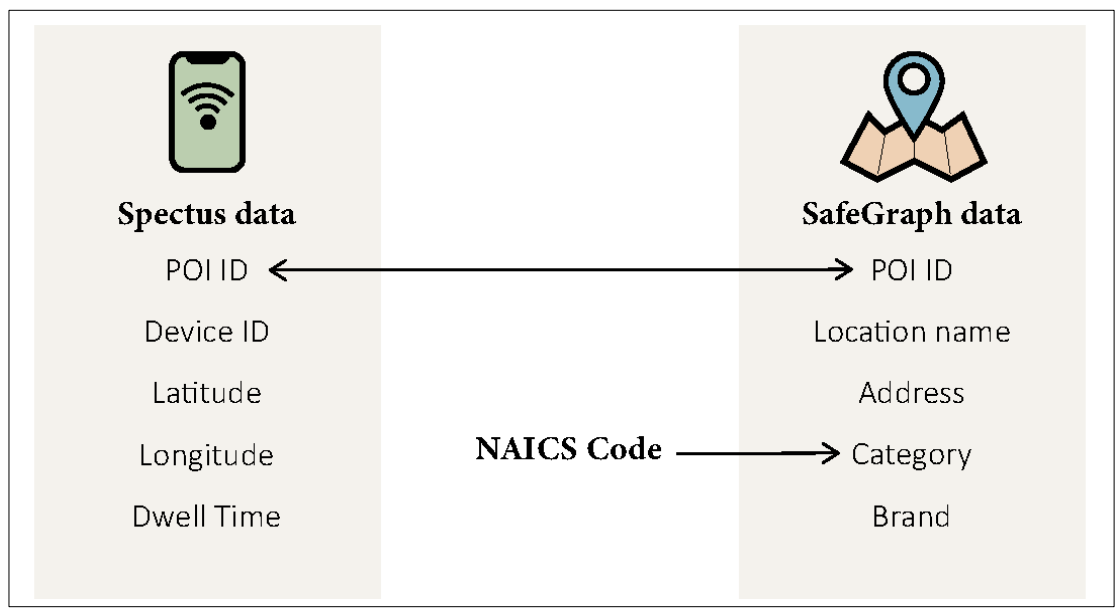

Fig 13 The data structure in this study

**Reference**

Alam, C. N., Manaf, K., Atmadja, A. R., & Aurum, D. K. (2016). Implementation of haversine formula for counting event visitor in the radius based on Android application. 2016 4th International Conference on Cyber and IT Service Management,

Albert, R., & Barabási, A.-L. (2002). Statistical mechanics of complex networks. *Reviews of Modern Physics*, *74*(1), 47-97. https://doi.org/10.1103/RevModPhys.74.47

Barrat, A., Barthelemy, M., Pastor-Satorras, R., & Vespignani, A. (2004). The architecture of complex weighted networks. *Proceedings of the National Academy of Sciences*, *101*(11), 3747-3752.

Batty, M. (2013). *The new science of cities*. MIT press.

Benson, A. R., Gleich, D. F., & Leskovec, J. (2016). Higher-order organization of complex networks. *Science*, *353*(6295), 163-166.

Cao, J., Li, Q., Tu, W., Gao, Q., Cao, R., & Zhong, C. (2021). Resolving urban mobility networks from individual travel graphs using massive-scale mobile phone tracking data. *Cities*, *110*, 103077.

Cao, J., Li, Q., Tu, W., & Wang, F. (2019). Characterizing preferred motif choices and distance impacts. *Plos one*, *14*(4), e0215242.

Dey, A. K., Gel, Y. R., & Poor, H. V. (2019). What network motifs tell us about resilience and reliability of complex networks. *Proceedings of the National Academy of Sciences*, *116*(39), 19368-19373. https://doi.org/doi:10.1073/pnas.1819529116

Di Clemente, R., Luengo-Oroz, M., Travizano, M., Xu, S., Vaitla, B., & González, M. C. (2018). Sequences of purchases in credit card data reveal lifestyles in urban populations. *Nature Communications*, *9*(1), 1-8.

González, M. C., Hidalgo, C. A., & Barabási, A.-L. (2008). Understanding individual human mobility patterns. *Nature*, *453*(7196), 779-782. https://doi.org/10.1038/nature06958

Hasnat, M. M., & Hasan, S. (2018). Identifying tourists and analyzing spatial patterns of their destinations from location-based social media data. *Transportation Research Part C: Emerging Technologies*, *96*, 38-54.

Huang, Q., & Wong, D. W. (2016). Activity patterns, socioeconomic status and urban spatial structure: what can social media data tell us? *International Journal of Geographical Information Science*, *30*(9), 1873-1898.

Jiang, S., Ferreira, J., & Gonzalez, M. C. (2017). Activity-Based Human Mobility Patterns Inferred from Mobile Phone Data: A Case Study of Singapore. *IEEE Transactions on Big Data*, *3*, 208-219. https://doi.org/10.1109/TBDATA.2016.2631141

Jurdak, R., Zhao, K., Liu, J., AbouJaoude, M., Cameron, M., & Newth, D. (2015). Understanding human mobility from Twitter. *Plos one*, *10*(7), e0131469.

Lazer, D., Pentland, A., Adamic, L., Aral, S., Barabási, A.-L., Brewer, D., Christakis, N., Contractor, N., Fowler, J., Gutmann, M., Jebara, T., King, G., Macy, M., Roy, D., & Alstyne, M. V. (2009). Computational Social Science. *Science*, *323*(5915), 721-723. https://doi.org/doi:10.1126/science.1167742

Lei, D., Chen, X., Cheng, L., Zhang, L., Ukkusuri, S. V., & Witlox, F. (2020). Inferring temporal motifs for travel pattern analysis using large scale smart card data. *Transportation Research Part C: Emerging Technologies*, *120*, 102810.

Lenormand, M., Louail, T., Cantú-Ros, O. G., Picornell, M., Herranz, R., Arias, J. M., Barthelemy, M.,

Miguel, M. S., & Ramasco, J. J. (2015). Influence of sociodemographic characteristics on human mobility. *Scientific reports*, *5*(1), 10075. https://doi.org/10.1038/srep10075

Li, Q., Bessell, L., Xiao, X., Fan, C., Gao, X., & Mostafavi, A. (2021). Disparate patterns of movements and visits to points of interest located in urban hotspots across US metropolitan cities during COVID-19. *Royal Society open science*, *8*(1), 201209.

Louail, T., Lenormand, M., Picornell, M., García Cantú, O., Herranz, R., Frias-Martinez, E., Ramasco, J. J., & Barthelemy, M. (2015). Uncovering the spatial structure of mobility networks. *Nature Communications*, *6*(1), 6007. https://doi.org/10.1038/ncomms7007

Maeda, T. N., Shiode, N., Zhong, C., Mori, J., & Sakimoto, T. (2019). Detecting and understanding urban changes through decomposing the numbers of visitors' arrivals using human mobility data. *Journal of Big Data*, *6*(1), 1-25.

McKenzie, G. (2020). Urban mobility in the sharing economy: A spatiotemporal comparison of shared mobility services. *Computers, Environment and Urban Systems*, *79*, 101418.

Milo, R., Shen-Orr, S., Itzkovitz, S., Kashtan, N., Chklovskii, D., & Alon, U. (2002). Network motifs: simple building blocks of complex networks. *Science*, *298*(5594), 824-827.

Moro, E., Calacci, D., Dong, X., & Pentland, A. (2021). Mobility patterns are associated with experienced income segregation in large US cities. *Nature Communications*, *12*(1), 1-10.

Newman, M. (2018). *Networks*. Oxford university press.

Opsahl, T., & Panzarasa, P. (2009). Clustering in weighted networks. *Social networks*, *31*(2), 155-163.

Saberi, M., Ghamami, M., Gu, Y., Shojaei, M. H. S., & Fishman, E. (2018). Understanding the impacts of a public transit disruption on bicycle sharing mobility patterns: A case of Tube strike in London. *Journal of transport geography*, *66*, 154-166.

Schneider, C., Belik, V., Couronné, T., Smoreda, Z., & Gonzalez, M. C. (2013). Unravelling daily human mobility motifs. *Journal of the Royal Society, Interface / the Royal Society*, *10*, 20130246. https://doi.org/10.1098/rsif.2013.0246

Schneider, C. M., Rudloff, C., Bauer, D., & González, M. C. (2013). Daily travel behavior: lessons from a week-long survey for the extraction of human mobility motifs related information. Proceedings of the 2nd ACM SIGKDD international workshop on urban computing,

Scott, J., & Carrington, P. J. (2011). *The SAGE handbook of social network analysis*. SAGE publications.

Song, C., Koren, T., Wang, P., & Barabási, A.-L. (2010). Modelling the scaling properties of human mobility. *Nature Physics*, *6*(10), 818-823. https://doi.org/10.1038/nphys1760

Song, C., Qu, Z., Blumm, N., & Barabási, A.-L. (2010). Limits of Predictability in Human Mobility. *Science*, *327*(5968), 1018-1021. https://doi.org/doi:10.1126/science.1177170

Stone, L., Simberloff, D., & Artzy-Randrup, Y. (2019). Network motifs and their origins. *PLoS computational biology*, *15*(4), e1006749.

Su, R., McBride, E. C., & Goulias, K. G. (2020). Pattern recognition of daily activity patterns using human mobility motifs and sequence analysis. *Transportation Research Part C: Emerging Technologies*, *120*, 102796. https://doi.org/https://doi.org/10.1016/j.trc.2020.102796

Toole, J. L., Herrera-Yaqüe, C., Schneider, C. M., & González, M. C. (2015). Coupling human mobility and social ties. *Journal of The Royal Society Interface*, *12*(105), 20141128.

Xu, S., Morse, S., & Gonzalez, M. C. (2020). *Modeling Human Dynamics and Lifestyle Using Digital Traces*.

Xu, Y., Shaw, S.-L., Zhao, Z., Yin, L., Fang, Z., & Li, Q. (2015). Understanding aggregate human mobility patterns using passive mobile phone location data: a home-based approach. *Transportation*, *42*(4), 625-646.

Yang, X., Fang, Z., Xu, Y., Yin, L., Li, J., & Lu, S. (2019). Spatial heterogeneity in spatial interaction of human movements—Insights from large-scale mobile positioning data. *Journal of transport geography*, *78*, 29-40.

Yang, Y., Heppenstall, A., Turner, A., & Comber, A. (2019). A spatiotemporal and graph-based analysis of dockless bike sharing patterns to understand urban flows over the last mile. *Computers, Environment and Urban Systems*, *77*, 101361.

Yin, J., & Chi, G. (2020). *Characterizing People's Daily Activity Patterns in the Urban Environment: A Mobility Network Approach with Geographic Context-Aware Twitter Data*.

Zhang, X., Xu, Y., Tu, W., & Ratti, C. (2018). Do different datasets tell the same story about urban mobility—A comparative study of public transit and taxi usage. *Journal of transport geography*, *70*, 78-90.

Zhang, Z., Long, Y., Chen, L., & Chen, C. (2021). Assessing personal exposure to urban greenery using wearable cameras and machine learning. *Cities*, *109*, 103006.

Zhao, C., Zeng, A., & Yeung, C. H. (2021). Characteristics of human mobility patterns revealed by high-frequency cell-phone position data. *EPJ Data Science*, *10*(1), 5.